\documentclass{ws-procs9x6}

\usepackage{graphicx}
\usepackage[normalem]{ulem}
\usepackage{color}
\definecolor{blue}{rgb}{0,0,1}
\definecolor{red}{rgb}{1,0,0}

\newcommand{\ecm}{\,e\mathrm{cm}}

\begin{document}

\title{CLARIFICATION ON THEORETICAL PREDICTIONS FOR GENERAL RELATIVISTIC EFFECTS IN FROZEN SPIN STORAGE RINGS}

\author{Andr\'as L\'aszl\'o, Zolt\'an Zimbor\'as}

\address{Wigner Research Centre for Physics,\\
P.O.Box 49, H-1525 Budapest, Hungary\\
\texttt{laszlo.andras@wigner.hu}\\
\texttt{zimboras.zoltan@wigner.hu}}

\begin{abstract}
Electromagnetic moments of particles carry important information on their 
internal structure, as well as on the structure of the effective Lagrangian 
describing their underlying field theory. One of the cleanest observable 
of such kind is the electric dipole moment (EDM), since Standard Model 
estimates would imply very small, much less then $10^{-30}\ecm$ 
value for that quantity, whereas several Beyond Standard Model 
(BSM) theories happen to predict of the order of $10^{-28}\ecm$ EDM 
for elementary or hadronic particles. 
So far, precision EDM upper bounds are mainly available for neutrons via cold 
neutron experiments, and indirect measurements for electrons. 
Therefore, in the recent year there has been a growing interest for direct measurement 
of EDM for charged particles, such as electrons, protons, muons or light nuclei. 
Such measurements become possible in relativistic storage rings, called 
\emph{frozen spin storage rings}. Many environmental factors give systematic 
backgrounds to the EDM signal, including General Relativity (GR), due to 
the gravitational field of the Earth. It turns out that, depending of the 
experimental scenario, the GR effect can be well above the planned EDM 
sensitivity. Therefore, it is both of concern as a source of systematics, 
as well as it can serve as a spin-off experiment for an independent test 
of GR. There are a handful of theoretical papers quantifying the GR systematics, 
delivering slightly different results. The aim of this paper is to clarify 
these claims, eventually try to reconcile these predictions, and to 
deduce their experimental implications. The closing section of the paper 
quantifies the field imperfection systematic error cancellation in 
the case of a so-called \emph{doubly-frozen spin storage ring} setting, 
in the idealized axial symmetric limit.
\end{abstract}


\vspace{5mm}

\bodymatter

\section{Introduction}
\label{introduction}

In the recent years, a significant interest built up for studying the 
electric dipole moment (EDM) of charged elementary or hadronic particles 
(electrons, muons, protons, or light nuclei). The reason for that is 
the following fact: the Standard Model (SM) predicts rather low, way smaller 
than $10^{-30}\ecm$ EDM, whereas several most prominent Beyond Standard Model 
(BSM) theories predict EDMs of the order of $10^{-28}\ecm$. As such, setting 
experimental upper bounds to EDM can serve as one of the most sensitive 
SM/BSM discriminator. So far, stringent EDM upper bounds were obtained 
for neutrons via ultracold neutron experiments, as well as indirect 
measurements for electrons were also obtained for bound electrons in large 
$Z$ atoms \cite{kirch2020}. For unbound charged particles, the measurement 
becomes obviously difficult, since their EDM cannot be simply be determined 
via putting them into a homogeneous electrostatic field and observing the 
transition between the parallel and antiparallel spin state to the electric 
field. Therefore, the technically rather challenging idea of \emph{frozen 
spin storage ring} emerged in the early 2000s (see reviews e.g.\ in \cite{semertzidis2016,talman2017}).

\begin{figure}[!h]
\includegraphics[width=9cm]{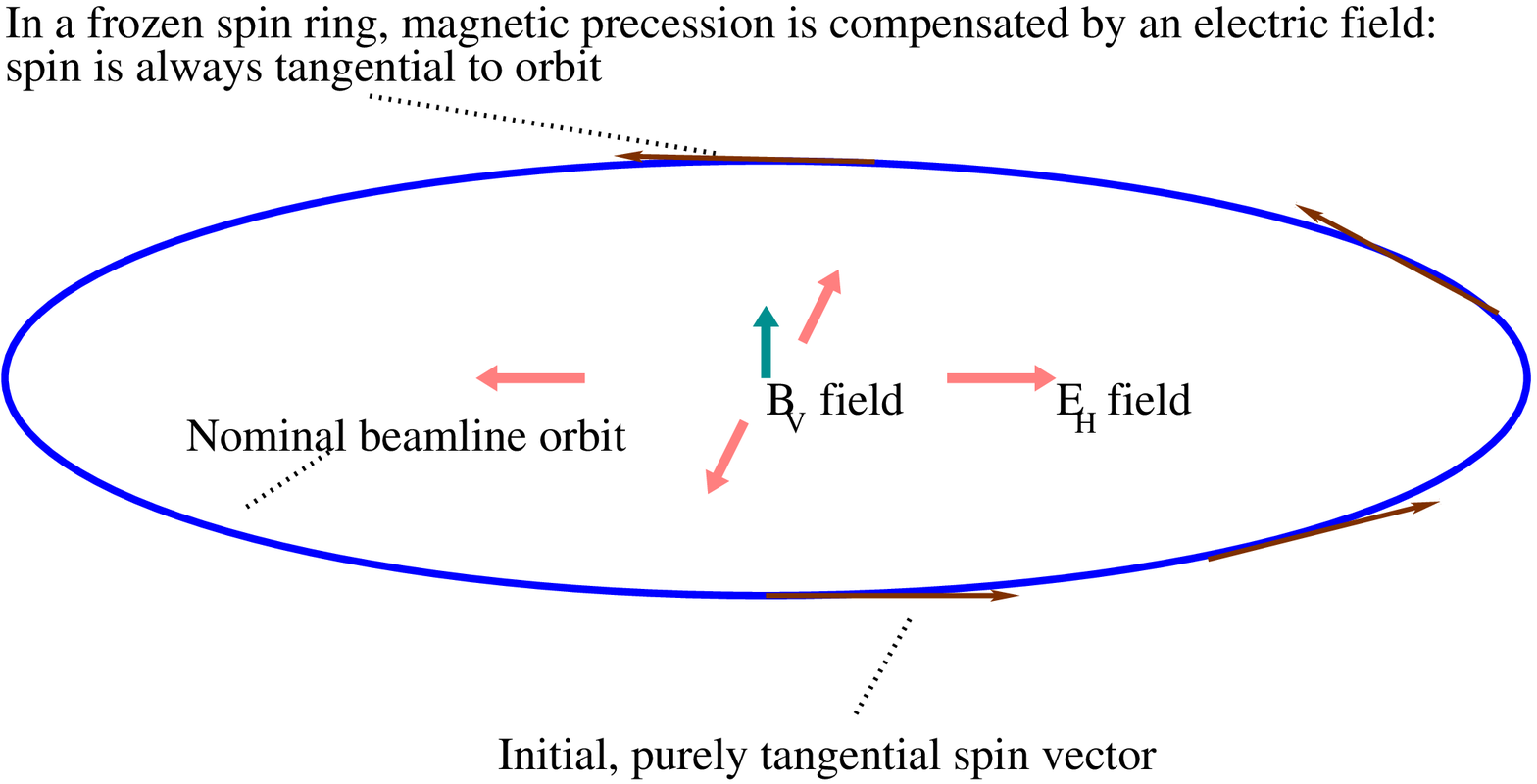}\hspace*{4mm}\begin{minipage}{2.1cm}\vspace*{-4.4cm}\includegraphics[width=2cm]{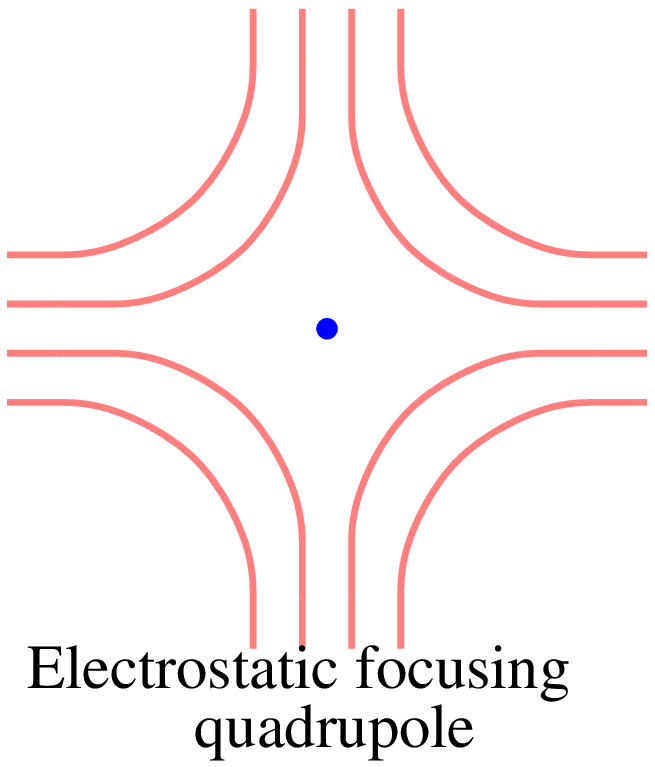}\end{minipage}
\caption{(Color online) Illustration of the concept of a frozen spin storage 
ring. The basic idea is that two kind of bending fields are used, a magnetic 
one and an electrostatic one, in such combination that both the circular motion 
is satisfied, as well as the magnetic precession of the spin vector is stalled. 
The transverse dispersion of the beam is kept under control by additional, 
electrostatic quadrupole beam focusing optics. For particles with $a>0$, 
it is possible to construct such rings with electrostatic-only bending fields, 
at the magic momentum, $\vert\beta\gamma\vert=\frac{1}{\sqrt{a}}$, 
where $\beta\gamma$ denotes momentum-over-mass.}
\label{figFS}
\end{figure}

In a typical idealized storage ring, particles are confined to a circular 
orbit using a homogeneous magnetic bending field. In such a situation, however, 
the spin of the particles precesses around the bending axis, proportionally 
to their magnetic moment anomaly, $a:=\frac{\mathsf{g}-2}{2}$. In a frozen 
spin storage ring, an additional, beam-radial electrostatic bending field 
is superimposed, such that the circular motion is satisfied, and the magnetic 
precession of the spin vector is stalled. This is illustrated in 
Figure~\ref{figFS}. In such condition, if the particle had an electric dipole 
moment, it would cause a spin precession around the instantaneous beam-radial axis.

The concept of frozen spin ring setting is suggested by the 
special relativistic equation of motion of point particles with spin, which 
are the Newton and the Thomas-Bargmann-Michel-Telegdi (TBMT) equations:
\begin{eqnarray}
 \frac{\mathrm{d}\vec{\beta}}{\mathrm{d}t_{{}_{\mathrm{lab}}}} \;=\; \frac{q}{m\,\gamma}\,\left(\vec{E}/c \;-\; (\vec{\beta}\cdot\vec{E}/c)\,\vec{\beta} \;+\; \vec{\beta}\times\vec{B}\right),\cr
  \cr
 \Bigg.\frac{\mathrm{d}\vec{S}_{{}_{\mathrm{lab,corot.}}}}{\mathrm{d}t_{{}_{\mathrm{lab}}}} \;=\; -\frac{q}{m}\cdot \qquad\qquad\qquad\qquad\qquad\qquad\qquad\; \cr
  \left(\underbrace{a\,\vec{B}}_{\substack{\text{magnetic}\\\text{term}}}+\underbrace{{\overbrace{\left(\frac{1}{\left(\beta\gamma\right)^{2}}-a\right)}^{\substack{\text{zero at}\\\text{``magic momentum''}}}}\vec{\beta}\times \vec{E}/c}_{\text{electric term}} + \underbrace{\frac{1}{2}\,\eta\left(\vec{E}/c+\vec{\beta}\times\vec{B}\right)}_{\text{EDM term}}\right) 
  \times\vec{S}_{{}_{\mathrm{lab,corot.}}} \Bigg.
\label{eqTBMTlab}
\end{eqnarray}
Here, $c$ is the speed of light, $m$ is the mass, $q$ is the charge of the particle, $s$ is its spin, 
$\mu$ is its magnetic dipole moment, $d$ is its electric dipole moment (EDM), 
$\mathsf{g}:=\frac{2\,m\,\mu}{q\,s}$ is its scaled magnetic moment, 
$a:=\frac{\mathsf{g}{-}2}{2}$ is its magnetic moment anomaly, 
$\eta:=\frac{2\,m\,c\,d}{q\,s}$ is its ``$\mathsf{g}$'' for the EDM. 
The fields $\vec{B}$, $\vec{E}$ are understood in the laboratory frame, 
as well as the time $t_{\mathrm{lab}}$, the velocity vector $\vec{\beta}$, 
and the spin vector $\vec{S}_{\mathrm{lab}}$. The vector 
$\vec{S}_{\mathrm{lab,corot.}}$ denotes the projections of the latter 
onto the tangent, normal and binormal directions, i.e.\ its components in the 
corotating (Fernet--Serret) coordinates. 
The frozen spin condition is said to be satisfied whenever 
$\frac{\mathrm{d}\vec{S}_{{}_{\mathrm{lab,corot.}}}}{\mathrm{d}t_{{}_{\mathrm{lab}}}}=0$ holds, 
assuming $\eta=0$. 
It is seen that for particles with $a>0$ one may set 
$\left\vert\beta\gamma\right\vert=\frac{1}{\sqrt{a}}$, 
which is also called the ``magic momentum'', and under such condition 
the electrostatic term does not contribute. At magic momentum, the frozen spin 
condition is satisfied whenever $\vec{B}=0$. More generally, at any 
momentum-over-mass $\beta\gamma$, the idealized planar circular motion together with 
the frozen spin condition is satisfied whenever
\begin{eqnarray}
E_{H}\cdot L & = & -\mathop{\mathrm{sign}}(a)\,\frac{m\,c^{2}}{q}\,\frac{(a\,\beta\gamma)^{2}\sqrt{a^{2}+(a\,\beta\gamma)^{2}}}{a^{2}\left(1+a\right)}, \cr
 & & \cr
B_{V}\cdot L & = & \qquad\qquad\;\, \frac{m\,c}{q}\;\;\frac{(a\,\beta\gamma)(a-(a\,\beta\gamma)^{2})}{a^{2}\left(1+a\right)}
\label{eqFr}
\end{eqnarray}
holds, where $L$ is the bending radius, $B_{V}$ is the homogeneous vertical 
magnetic bending field, and $E_{H}$ is the horizontal (beam-radial) electrostatic 
bending field, both sampled at the beam trajectory.

The experimental proposals for EDM rings \cite{abusaif2018} aim to reach an 
EDM sensitivity of the order of $10^{-29}\ecm$, which would be equivalent (in a typical 
realistic setting) to an instrumental sensitivity for about 
$10^{-9}\,\mathrm{rad/s}$ for the rate of spin precession around the 
instantaneous beam-radial axis. Technically, this is measured by initially 
longitudinally polarized beams, and this small precession rate is detected 
via the rate of vertical polarization buildup. This small signal adds up coherently 
with each full revolution, and therefore can be accumulated. In the foreseen 
experimental setting, the signal integration time is of the order of an hour.

\section{General Relativistic (GR) effects}

In the years of early 2000s it was suggested \cite{silenko2007} that in 
storage ring experiments, Earth's gravitational field might cause a 
systematic effect on the spin precession. In manifestly covariant 
General Relativistic (GR) formalism, the Newton plus TBMT equations 
\cite{conte1996,jackson1999,hawking1973,wald1984} read as:
\begin{eqnarray}
  u^{a}\,\nabla_{a}u^{b} & = & -\frac{q}{m}\,g^{ab}\,F_{ac}\,u^{c}, \cr
 & & \cr
  D^{F}_{u}w^{b}         & = & -\frac{\mu}{s}\, \left(g^{ab}\,F_{ac} \,-\, u^{b}\,u^{d}\,F_{dc} \,-\, g^{ab}\,F_{ad}\,u^{d}\,u^{e}\,g_{ec}\right) \,w^{c},\cr
 \Bigg. & &                     +\frac{d}{s}\, \left(g^{ab}\,{}^{\star}\!\!F_{ac} \,-\, u^{b}\,u^{d}\,{}^{\star}\!\!F_{dc} \,-\, g^{ab}\,{}^{\star}\!\!F_{ad}\,u^{d}\,u^{e}\,g_{ec}\right) \,w^{c}.
\label{eqnt}
\end{eqnarray}
Here, $u^{a}$ is the four velocity vector field along the particle worldline, 
whereas $w^{a}$ denotes the spin direction vector, the 
symbol $F_{bc}$ denotes the electromagnetic 
field strength tensor of the total guiding fields, ${}^{\star}\!\!F_{bc}$ denotes the 
Hodge dual of the electromagnetic field strength tensor, $g_{ab}$ denotes the 
spacetime metric tensor field as usual, $\nabla_{a}$ denotes the spacetime 
covariant derivation compatible with the metric, and $D^{F}_{u}w^{b}$ denotes 
the \emph{Fermi--Walker derivative} of $w^{b}$ along the worldline described 
by $u^{a}$, and it is defined as 
$D^{F}_{u}w^{b}:=u^{a}\,\nabla_{a}w^{b}+g_{ed}w^{e}u^{b}u^{a}\nabla_{a}u^{d}-g_{cd}w^{c}u^{d}u^{a}\nabla_{a}u^{b}$. 
The Fermi--Walker derivative is a minimally modified version of the spacetime 
covariant derivation $\nabla_{a}$, preserving angles determined by the spacetime metric 
$g_{ab}$. As such, it can be interpreted as parallel transport of rigid frames 
along a prescribed timelike trajectory described by $u^{a}$. The rationale behind this equation 
of motion is that it preserves the orthogonality relation 
$g_{ab}u^{a}w^{b}=0$ of the four-velocity vector and the spin direction vector, 
which algebraic relation derives from the quantum mechanical origin of 
the spin vector (Pauli--Lyubanski vector). The pertinent algebraic constraint 
already causes a spin precession in special relativity, i.e.\ over Minkowski 
spacetime, whenever the trajectory described by $u^{a}$ is accelerating, 
i.e.\ $u^{a}\nabla_{a}u^{b}\neq 0$. That is called Thomas precession. 
For non-accelerating (geodesic) trajectories, 
the Fermi--Walker derivation falls back to ordinary covariant derivation, 
and Thomas precession is not present. 
Over general relativistic spacetimes, the precession caused by a Schwarzschild 
background is called the de Sitter precession or geodetic effect, 
and on a Kerr background it is called Lense--Thirring effect 
(also experimentally confirmed by the Gravity Probe B satellite experiment).

\section{Quantitative predictions of GR effects in EDM rings}

We now enumerate the available literature on the predictions for the 
GR systematics in a frozen spin ring EDM observable.

The work presented in \cite{silenko2007} seems to be the first one which 
suggests that in storage ring experiments, GR might have a contribution to 
the spin dynamics. In that work, a perturbative weak field approximation 
in terms of $\frac{r_{S}}{R}$ was used ($R$ standing for the Earth's radius, and 
$r_{S}$ standing for the Earth's Schwarzschild radius), 
worked out in laboratory frame formalism, i.e.\ in a non-manifestly 
covariant formalism. There was no concrete prediction stated, yet, 
for a frozen spin EDM ring.

The first concrete quantitative prediction for a frozen spin storage 
ring setting was presented in \cite{orlov2012}, but solely 
electrostatic-only (magic momentum) ring was considered. The authors determine 
that the total GR systematics in an electrostatic-only (magic momentum) 
frozen spin ring is
\begin{eqnarray}
\Omega_{GR}\Big\vert_{\text{magic momentum}} & = & -\sqrt{a}\, g/c,
\label{eqOrlov}
\end{eqnarray}
where $g:=\frac{r_{S}c^{2}}{2R^2}$ 
is the gravitational acceleration at the surface of the Earth. The result 
was obtained with a manifestly covariant formalism, in the weak field 
approximation, i.e.\ with a perturbative calculation in $\frac{r_{S}}{R}$. 
Since $g/c\approx 33\,\mathrm{nrad/s}$, this should be a significant 
background for an EDM experiment at the design sensitivity of 
$10^{-9}\,\mathrm{rad/s}$. 
One should recall that such an electric-only ring is only possible for particles with 
$a>0$ and at $\vert\beta\gamma\vert=\frac{1}{\sqrt{a}}$, i.e.\ such a 
setting cannot exist for the experimentally very relevant deuteron or helium-3 
nucleus \cite{stone2005} beams.

The work \cite{obukov2016,obukov2016b} presents the first concrete prediction 
for a mixed magnetic and electrostatic frozen spin storage ring, with 
an electrostatic or magnetostatic or combined focusing. The importance of this 
result is given by the above mentioned fact: such experimental setting 
can exist for all particle types and at any momenta, only the condition Eq.(\ref{eqFr}) needs 
to be satisfied (which is the frozen spin condition for an ideal planar circular ring).
The pertinent work used a perturbative 
approach in lab frame formalism, i.e. not a manifestly covariant formalism. 
The prediction given by the authors for GR systematics in a frozen spin ring is
\begin{eqnarray}
 \Omega_{GR} & = & \beta\, (1-a(2\gamma^2-1))/\gamma\, g/c
\label{eqObukov}
\end{eqnarray}
for the most important case of the electrostatic focusing (we only 
consider this case in this paper, for briefness).

The paper \cite{kobach2016} also discusses GR effect in storage rings, 
and draws a similar conclusion to \cite{silenko2007}. It does not give, 
however, a concrete prediction for an EDM ring. Similarly to 
\cite{silenko2007}, it uses perturbative approach in lab frame formalism, 
i.e.\ not in manifestly covariant formalism.

In the paper \cite{laszlo2018} again a quantitative prediction is derived 
for the GR systematics. For a mixed magnetic and electrostatic EDM ring, 
with electrostatic focusing, the prediction
\begin{eqnarray}
 \Omega_{GR} & = & -a\, \beta\gamma\, g/c
\label{eqLZ}
\end{eqnarray}
is derived. 
That paper intends to make a comprehensive, spacetime geometrical modeling 
of the idealized experimental setting over curved background. 
For cross-checking purposes, in that paper manifestly covariant formalism 
was used, and an exact solution was derived, without intermediary weak field 
perturbation approach in terms of $\frac{r_{S}}{R}$. As such, this prediction 
should be compared to that of \cite{obukov2016} (or to \cite{orlov2012} at 
the magic momentum setting).

\section{Discussion}

It is seen that the quantitative predictions in the literature 
\cite{orlov2012,obukov2016,laszlo2018} all agree for the electrostatic-only 
(magic momentum) ring case, and the according prediction for GR systematics is 
Eq.(\ref{eqOrlov}).

\begin{figure}[!h]
\begin{center}
\includegraphics[width=5.5cm]{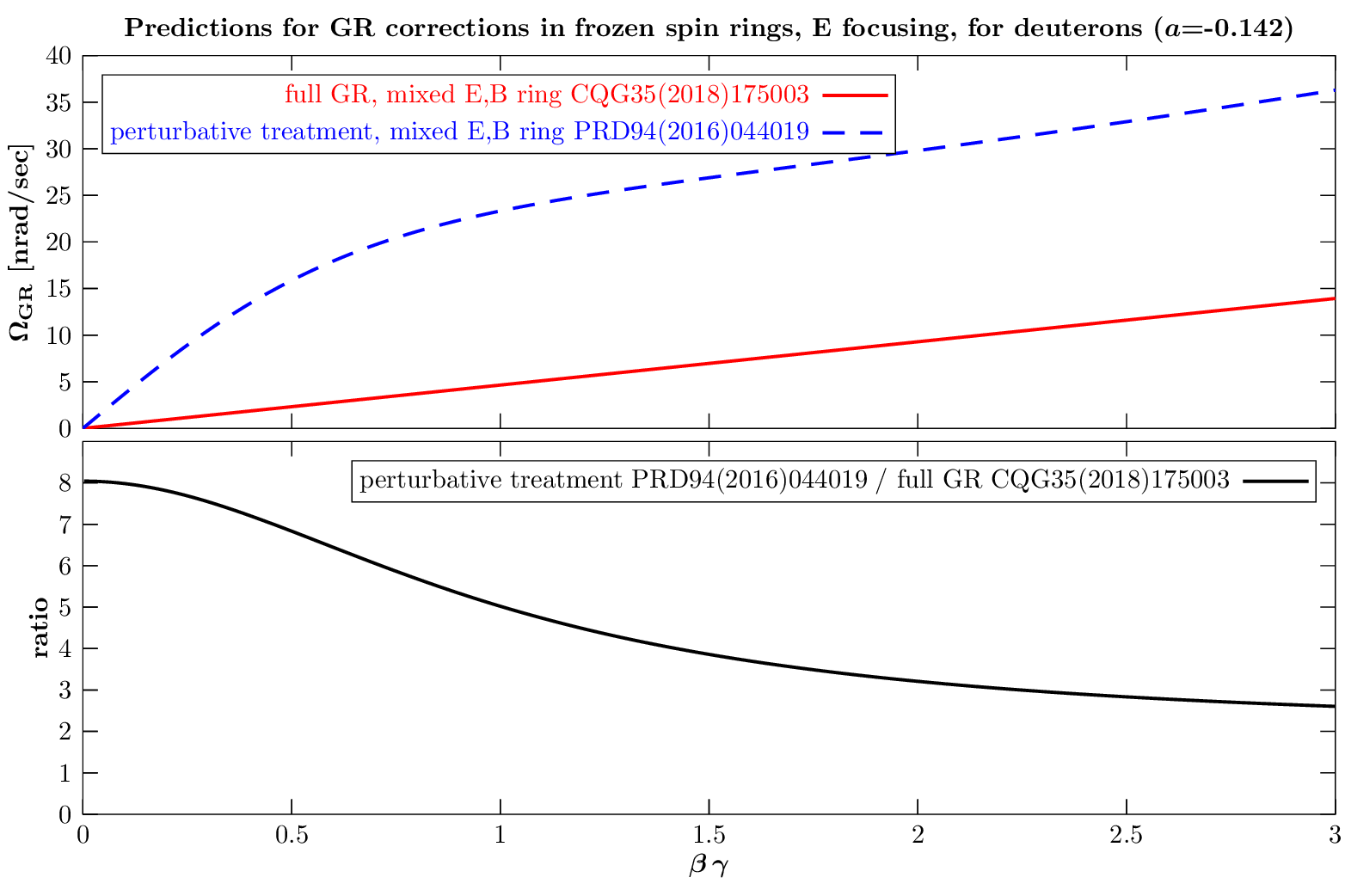}$\;$\includegraphics[width=5.5cm]{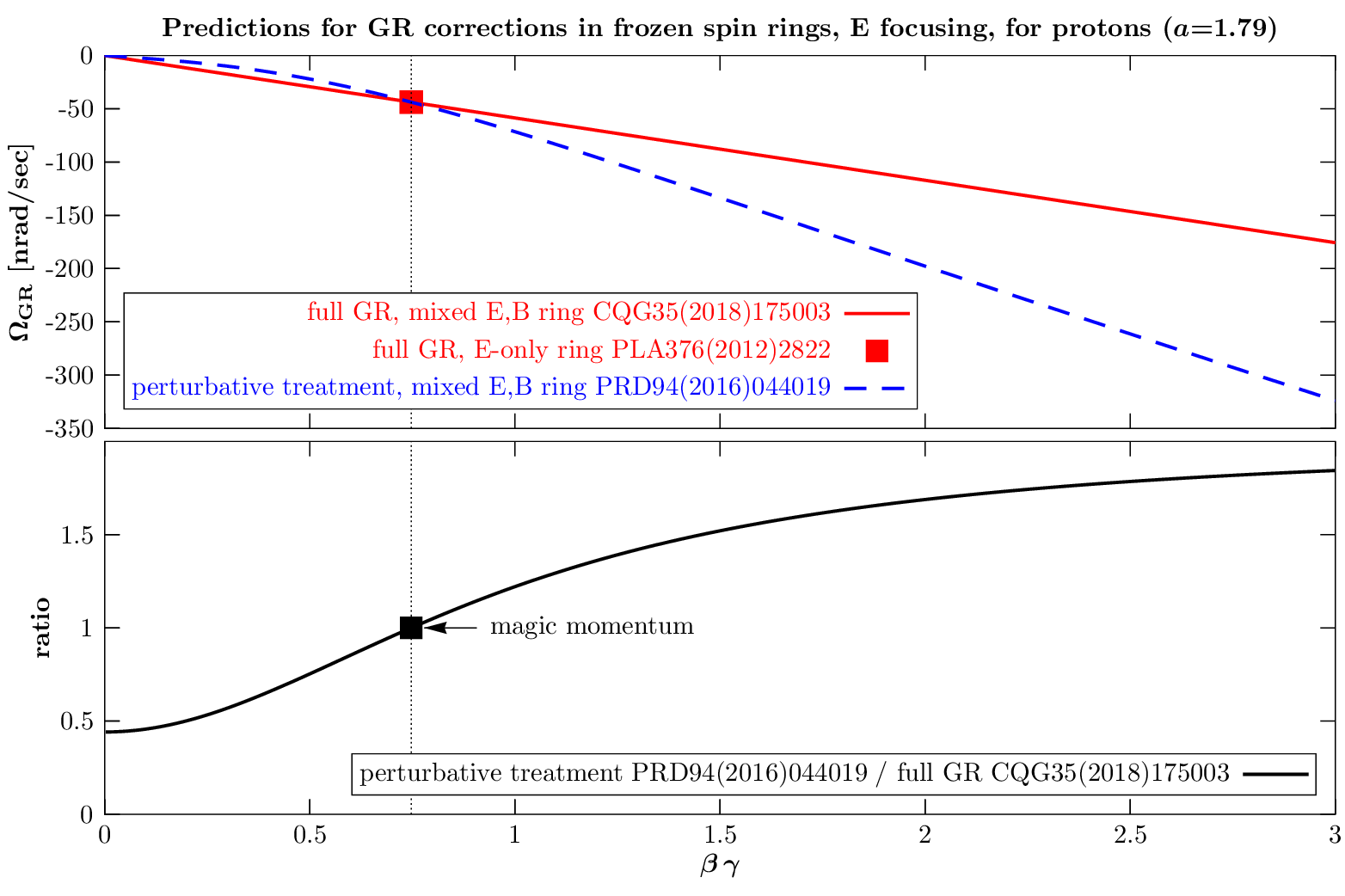}
\end{center}
\caption{(Color online) Comparison of predictions 
\cite{obukov2016} (Eq.(\ref{eqObukov})) and \cite{laszlo2018} (Eq.(\ref{eqLZ})) 
for the GR systematics $\Omega_{GR}$ in the EDM observable, for the case 
of the experimentally very relevant deuteron beam ($a\approx-0.142$, left panels) and 
proton beam ($a\approx1.79$, right panels). At the ``magic momentum'' (only possible 
for particles with $a>0$), all the predictions agree.}
\label{figDeu}
\end{figure}

From the experimental point of view, however, it is important to consider 
non-magic momentum rings as well. Apparently, for this general case, the 
predictions of \cite{obukov2016} and \cite{laszlo2018} 
(i.e.\ Eq.(\ref{eqObukov}) and Eq.(\ref{eqLZ})) differ, up to a factor 
of $2-8$ in the experimentally relevant $\beta\gamma$ and $a$ settings. 
The aim of the present paper is to identify and understand the source of 
this difference in the predictions. The relevance of these investigations 
is given by the fact that for the experimentally most relevant particle 
type (deuteron) one has $a\approx -0.142$, for which Eq.(\ref{eqObukov}) 
predicts a measurable effect well above the planned experimental sensitivity 
$\approx 1\,\mathrm{nrad/s}$, whereas Eq.(\ref{eqLZ}) 
would predict an effect below or just around the sensitivity, 
i.e.\ it foresees practically no observable effect. For protons, away from 
the magic momentum, the predictions also differ up to factor of $2$ in the 
extreme case. This comparison is visualized in Figure~\ref{figDeu}.

\begin{figure}[!h]
\includegraphics[width=9cm]{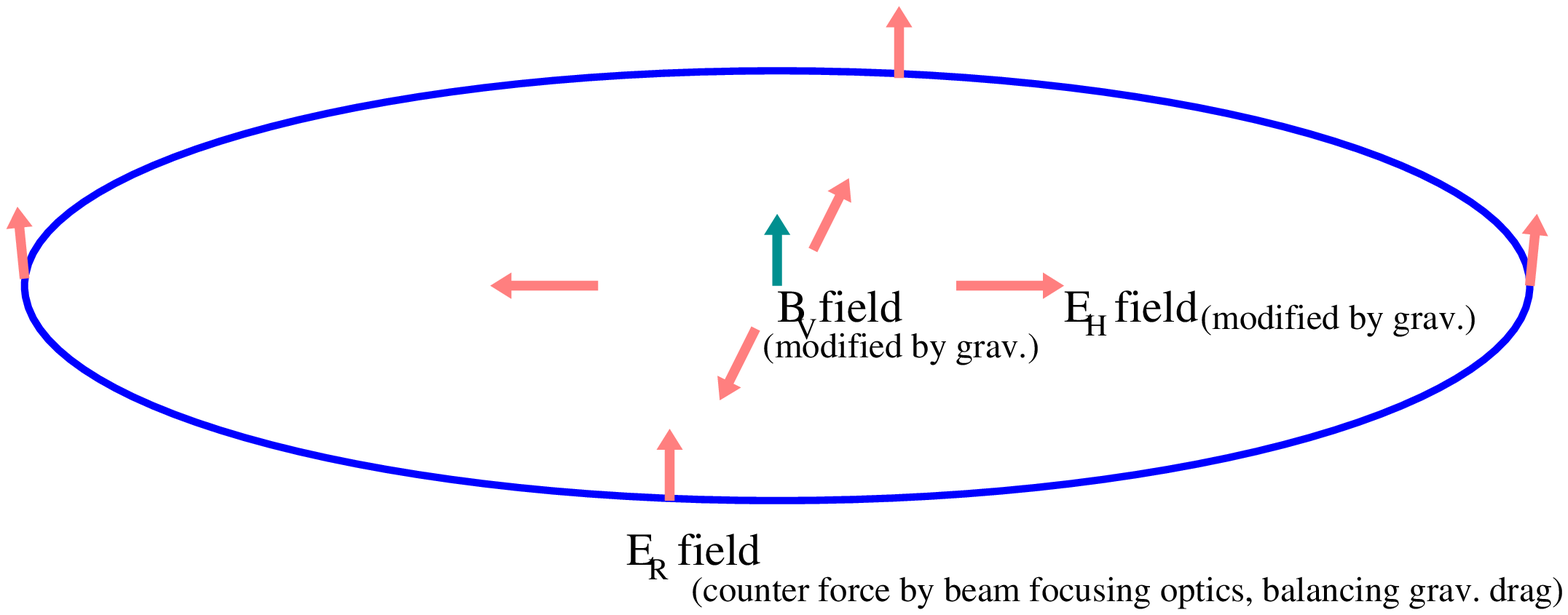}\hspace*{4mm}\begin{minipage}{2.1cm}\vspace*{-4.4cm}\includegraphics[width=2cm]{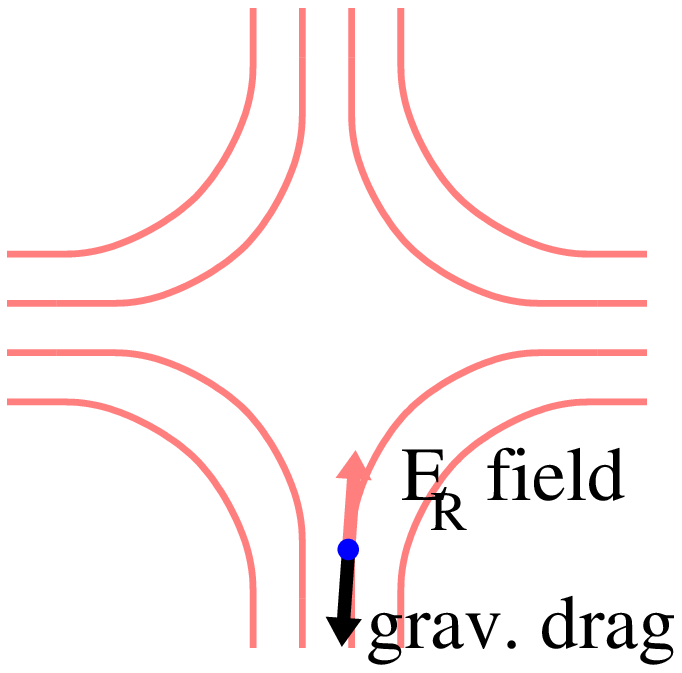}\end{minipage}
\caption{(Color online) Illustration of the GR modification effect of the 
guiding fields in a frozen spin ring. The magnetic bending field ($B_{V}$) 
is the asymptotically homogeneous field over Schwarzschild, the electric 
bending field ($E_{H}$) is the field of an infinite charged wire, and 
the focusing quadrupoles exert an Earth-radial ($E_{R}$) field to balance 
gravitational drag, at the position of the nominal planar circular closed 
orbit. The precise formulae for these over Schwarzschild 
is described in \cite{laszlo2018}.}
\label{figFRg}
\end{figure}

The main complication for the quantitative calculation of $\Omega_{GR}$ comes 
from the fact, that in general relativity, there is no model for a homogeneous 
gravitational field, which is valid globally. Therefore, one cannot just take 
e.g.\ a Minkowski limit of Eq.(\ref{eqnt}), and apply a homogeneous 
acceleration field, like one would be able to do in nonrelativistic mechanics. 
One therefore cannot avoid to put the equations of motion Eq.(\ref{eqnt}) on a 
Schwarzschild spacetime, and do the calculation in GR. 
Detailed analysis of the perturbative lab-frame formalism 
(i.e.\ non-manifestly covariant) calculation \cite{obukov2016} shows that in their 
calculation an implicit assumption was made, namely that the magnetic bending 
axis is Earth-radial. One should recall that Schwarzschild metric is modeling 
the round Earth and not an infinitely large Earth (which would stand for a 
homogeneous field), and 
in a perturbative approach in fact one needs to take care about \emph{two} 
small parameters: $\frac{r_{S}}{R}$ and $\frac{L}{R}$, the symbol $L$ 
as before standing for the storage ring radius. In an experimental situation 
at the surface of the Earth with a storage ring of bending radius $L\approx 10\,\mathrm{m}$, 
one has that $\frac{r_{S}}{R}\approx1.41\cdot10^{-9}$ and $\frac{L}{R}\approx1.57\cdot10^{-6}$. 
As such, if one wishes to keep track of the gravitational modification effects 
(perturbatively scaling with $\frac{r_{S}}{R}$), one cannot just disregard 
the effect coming from the curvature of the Earth 
(perturbatively scaling with $\frac{L}{R}$): one cannot simply approximately 
equate the Earth-radial direction with the ``vertical'' direction (being 
the ring axis direction). To model the magnetic bending fields correctly, 
one actually needs to calculate the asymptotically homogeneous magnetic 
field over Schwarzschild spacetime (Eq.(53) and Figure~3 in \cite{laszlo2018}). 
To model the electrostatic bending field, which is the electric field of an 
infinite homogeneously charged wire, one needs to calculate that field 
over Schwarzschild (Eq.(59) and Figure~4 in \cite{laszlo2018}). 
Quite naturally, the beam needs to be balanced against the Earth's gravitational 
drag. In the real experimental setting, as a consequence of the gravitational 
drag, the beam sinks into the electrostatic quadrupole focusing field, as 
depicted in Figure~\ref{figFRg}. 
Therefore, at the nominal 
equilibrium trajectory, an Earth-radial electrostatic field will balance 
the gravitational drag (Eq.(55) and Figure~3 in \cite{laszlo2018}).

\begin{figure}[!h]
\begin{center}
\includegraphics[width=2cm]{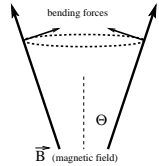}
\end{center}
\caption{(Color online) If the magnetic bending field is not vertical, 
but e.g.\ Earth-radial, then a small beam-radial $B_{H}$ component will be present besides $B_{V}$. 
The component $B_{H}$ will first of all induce a small vertical Lorentz force which is superimposed 
onto the gravitational drag, and keeps equilibrium with the force of the 
quadrupole focusing optics. 
Furthermore, the small beam-radial component $B_{H}$ acts with a direct torque 
on the spin, which is a large effect.}
\label{figB}
\end{figure}

In the model, the amplitudes of the above three guiding fields $B_{V}$, $E_{H}$, 
$E_{R}$ are uniquely 
determined by the condition that the stationary planar circular motion 
holds, and that the horizontal component of the spin vector is frozen. 
If, on the contrary, one assumes an Earth-radial magnetic bending field, 
see Figure~\ref{figB}, 
then one can also satisfy the above condition, but at the limit of 
$r_{S}\rightarrow0$ the spin will still precess around the instantaneous 
beam axis due to the conical shape of the magnetic bending field 
(since the limit $r_{S}\rightarrow0$ does not imply $L\rightarrow0$).

Indeed, concrete calculation shows using the non-perturbative manifestly 
covariant formalism of \cite{laszlo2018}, that if the magnetic bending 
field at the nominal beamline is fixed to be Earth-radial direction, and a stationary 
closed planar circular nominal beam orbit as well as a frozen horizontal 
spin condition is assumed, then the spin precession rate around the 
instantaneous beam axis becomes:
\begin{eqnarray}
 \delta\!\Omega & = & \underbrace{\frac{c}{R}\,\beta\,\frac{a\,\beta^{2}\gamma^{2}-1}{\gamma}}_{\substack{\text{from Earth-radial conicality}\\\text{of bending fields}}} \;+\; \underbrace{\frac{1-a\,(2\gamma^{2}-1)}{\gamma}\,\beta\,c\,\frac{r_{S}}{2R^{2}}}_{\substack{\text{first order GR correction,}\\\text{PRD94(2016)044019}}} \;+\; O(r_{S}{}^{2}).
\label{eqOmegaR}
\end{eqnarray}
Consequently, one can read off from Eq.(\ref{eqOmegaR}), that the result of \cite{obukov2016} or 
Eq.(\ref{eqObukov}) is indeed correct as the first order GR correction, but 
for an Earth-radial magnetic bending field assumption. From the experimental 
point of view, it is important to note, that 
in such setting, even when GR is neglected (${r_{S}\rightarrow0}$), 
there is a rather large residual systematic contribution to the EDM 
observable, as seen from Eq.(\ref{eqOmegaR}). This is simply because of the conical shape 
imperfection of the magnetic bending field due to the Earth-radiality assumption, 
which will not vanish in the Minkowski limit. 
As such, the GR correction quantified in \cite{obukov2016} rather applies to 
a special case of a so-called \emph{Koop spin wheel} \cite{Mane2015} experiment. 
(Koop spin wheel is a modified version of a frozen spin setting, when bending 
and focusing fields are adjusted such that the planar circular motion is 
satisfied, and the horizontal spin is frozen, but the spin vector is allowed 
to precess slowly around the instantaneous beam-radial axis with a controllable 
rate. In this case, the EDM or GR signal adds to the spin roll coherently.)

The above claim can be explicitly shown of course by lengthy but rather straightforward 
calculations of directly solving Eq.(\ref{eqnt}) over Schwarzschild, 
similarly as done in \cite{laszlo2018}. It can, however, be justified by somewhat 
simpler means as well. For instance an intermediary result of \cite{obukov2016} 
is that in a magnetic-electric frozen spin ring with electrostatic focusing, 
the amplitude of the Earth-radial electric field exerted by the focusing 
quadrupoles is $q\,E_{R}=m\,\frac{2\gamma^{2}-1}{\gamma}\,g$ at equilibrium. This formula, 
at a first glance is quite striking, as it contradicts the naive expectation 
$m\,\gamma\,g$ from equivalence principle. In fact, with a vertical magnetic 
bending field and corresponding electrostatic bending field one can assess that 
in that situation rather $q\,E_{R}=m\,\gamma\,g$ holds. We show this below, 
explicitly.

For the calculations, let us use Schwarzschild coordinates $t,r,\vartheta,\varphi$ 
in a standard Schwarzschild spacetime, and let us use $c=1$ units for 
the remaining of the paper. An idealized planar circular closed 
stationary beam orbit on the surface of the Earth is an $r=const$ and 
$\vartheta=const$ worldline. The corresponding constants will be denoted by 
$R$ and $\Theta$. One can set $L:=R\sin\Theta$ for 
the beam bending radius, so then one has $\frac{L}{R}=\sin\Theta$ by definition. 
Such a world line has four-acceleration vector
\begin{eqnarray}
 u^{a}\nabla_{a}u^{b} & = & \underbrace{\frac{\mathrm{d}u^{b}}{\mathrm{d}\tau}}_{ = 0} + u^{a}u^{c}\Gamma_{ac}^{b} \;=\; \left(\begin{array}{c} \Big.0 \cr \Big.-\beta^{2}\gamma^{2}\frac{1}{R}\left(1-\frac{r_{S}}{R}\right)+\gamma^{2}\frac{r_{S}}{2R^{2}} \cr \Big.-\beta^{2}\gamma^{2}\frac{1}{L}\frac{1}{R}\sqrt{1-\frac{L^{2}}{R^{2}}} \cr \Big.0 \end{array}\right).
\label{eqa}
\end{eqnarray}
From this, as also pointed out in \cite{obukov2016,obukov2016b}, one may obtain 
that the GR correction to the Earth-radial projection of the four-acceleration 
vector is:
\begin{eqnarray}
 -g_{ab}\,\hat{r}^{a}\;(u^{c}\nabla_{c}u^{b}) & = & \!\!\!\!\!\!\!\!\underbrace{-\gamma^{2}\frac{\beta^{2}}{L}\frac{L}{R}}_{\substack{\text{constant part}\\\text{(Earth-radial projection of}\\\text{centrifugal four-acceleration}\\\text{in Minkowski limit)}}} \;+\; \underbrace{(2\gamma^{2}-1)\frac{r_{S}}{2R^{2}}}_{\text{first order GR correction}} \;+\; O(r_{S}{}^{2}).\cr
 & & 
\label{eqar}
\end{eqnarray}
The constant part is just the Earth-radial projection of the centrifugal 
four-acceleration vector in the Minkowski limit, and the first order term 
in $r_{S}$ is the first order GR correction. One would naively draw the conclusion 
from this, that the necessary Earth-radial electrostatic force needed 
to compensate for the gravitational drag is $q\,E_{R}=m\,\frac{2\gamma^{2}-1}{\gamma}\,g$, 
as also concluded e.g.\ in \cite{obukov2016}. This formula 
is apparently not in accordance with a naive application of the equivalence 
principle, which would dictate $q\,E_{R}=m\,\gamma\,g$. One should note, 
however, that not only the four-acceleration in the Newton equation gets a 
GR correction, but also the four-force expression of the electromagnetic 
forces. With vertical magnetic bending axis, as discussed and used in 
\cite{laszlo2018}, the vector of four-force over mass is:
\begin{eqnarray}
 -\frac{q}{m}\,g^{bc}\,F_{cd}\,u^{d} & = & -\frac{q}{m}\,\left(\begin{array}{c} \Big.0 \cr \Big.\frac{\left(B_{V}\,\beta\gamma-E_{H}\,\gamma\right)\,\frac{L}{R}\,\left(1-\frac{r_{S}}{R}\right)}{\sqrt{1-\frac{r_{S}}{R}\frac{L^{2}}{R^{2}}}}-E_{R}\,\gamma\,\sqrt{1-\frac{r_{S}}{R}} \cr \Big.\frac{1}{R}\frac{\left(B_{V}\,\beta\gamma-E_{H}\,\gamma\right)\sqrt{1-\frac{L^{2}}{R^{2}}}}{\sqrt{1-\frac{r_{S}}{R}\frac{L^{2}}{R^{2}}}} \cr \Big.0 \end{array}\right)
\label{eqf}
\end{eqnarray}
which needs to be equal to Eq.(\ref{eqa}) for the four-Newton equation to 
be satisfied. Apparently, this equation is triangular and therefore may 
be solved exactly. One infers:
\begin{eqnarray}
 q\,E_{R} & = & \underbrace{0}_{\substack{\text{zero constant offset}\\\text{in Minkowski limit}}} + \underbrace{m\gamma\, \frac{r_{S}}{2R^{2}}}_{\substack{\text{first order GR correction,}\\\text{as in CQG35(2018)175003}}} +\; O(r_{S}{}^{2})
\end{eqnarray}
as expected from naive application of the equivalence principle. 
If one, however, assumes an Earth-radial magnetic bending field (its amplitude 
denoted by $B$), then the expression for the vector of four-force over mass 
vector is:
\begin{eqnarray}
 -\frac{q}{m}\,g^{bc}\,F_{cd}\,u^{d} & = & -\frac{q}{m}\,\left(\begin{array}{c} \Big.0 \cr \Big.-E_{R}\,\gamma\,\sqrt{1-\frac{r_{S}}{R}} \cr \Big.\frac{1}{R}\left(B\,\beta\gamma-E_{H}\,\gamma\right) \cr \Big.0 \end{array}\right)
\end{eqnarray}
in this case. The four-Newton equation is satisfied whenever this equals to 
Eq.(\ref{eqa}). That also can be solved exactly, since it is a triangular 
equation, and one infers:
\begin{eqnarray}
 q\,E_{R} & = & \underbrace{-m\gamma\,\frac{\beta^{2}}{L}\,\frac{L}{R}}_{\substack{\text{a constant offset}\\\text{in Minkowski limit}\\\text{from Earth-radial bending axis}}} + \underbrace{m\,\frac{2\gamma^{2}-1}{\gamma}\,\frac{r_{S}}{2R^{2}}}_{\substack{\text{first order GR correction,}\\\text{as in PRD94(2016)044019}}} +\; O(r_{S}{}^{2})
\end{eqnarray}
where one can recognize that there is a constant offset due to the Earth-radial 
projection of the bending forces in Minkowski limit (see also the illustration 
Figure~\ref{figB}), and the GR correction is just as claimed in \cite{obukov2016}. 
According to the non-perturbative and manifestly covariant calculation, this 
should be understood to be in pair with Eq.(\ref{eqOmegaR}).

One can also reconcile the results of \cite{obukov2016} and \cite{laszlo2018} 
via allowing for four types of guiding fields: $B_{V}$, $E_{H}$, $E_{R}$ as 
previously, and an additional beam-radial magnetic field component $B_{H}$ at 
the nominal stationary planar circular beam line. Assuming that the planar 
circular motion is satisfied, as well as the horizontal spin component is 
frozen, then as a function of the freely specifiable parameters $\beta\gamma$ 
and $B_{H}$, the field amplitudes $B_{V}$, $E_{H}$ and $E_{R}$ are uniquely 
determined, moreover one gets a precession rate \cite{laszlo2019}
\begin{eqnarray}
 \delta\!\Omega & = & \underbrace{-\frac{q\,(1+a)}{m}\frac{1}{\gamma^{2}}\,B_{H}}_{\substack{\text{magnetic field conicality}\\\text{imperfection term}}} \quad+\quad \underbrace{-a\,\beta\gamma\;\frac{r_{S}}{2R^{2}}}_{\text{first order GR term}} \;+\; O(r_{S}{}^{2})
\label{eqconical}
\end{eqnarray}
around the instantaneous beam-radial axis (this would give a coherently 
accumulating background to the EDM signal). We derived this formula again 
with the same non-perturbative manifestly covariant formalism as used 
in \cite{laszlo2018}. The magnetic field conicality shape imperfection term 
(the first term) in this equation can also be derived, of course, without considering GR, 
namely from the Minkowski limit Newton + TBMT equations in the lab frame, i.e.\ merely from Eq.(\ref{eqTBMTlab}). 
Knowing this result, one can recover the GR correction of \cite{obukov2016}, 
i.e.\ Eq.(\ref{eqObukov}), or more precisely Eq.(\ref{eqOmegaR}). Namely, 
a fixed Earth-radial magnetic bending field would imply a conical field shape 
imperfection characterized by $\frac{B_{H}}{B_{V}}=\tan(\Theta)\sqrt{1-\frac{r_{S}}{R}}$. 
Plugging this identity into Eq.(\ref{eqconical}), one recovers Eq.(\ref{eqOmegaR}), 
using the Newton equation and the frozen horizontal spin condition. 
One can thus conclude that the quantitative difference between the 
prediction of \cite{obukov2016} and \cite{laszlo2018} can be well understood.

\section{Systematic error reduction using doubly-frozen spin ring}

The above discussed apparent discrepancy also highlights a rather important 
experimental fact, well-known by the EDM experimental community. Namely, 
it follows from the first term of Eq.(\ref{eqconical}), that a conical 
magnetic bending field shape imperfection (i.e.\ a small stray beam-radial magnetic 
field component) gives substantial contribution to the EDM observable. 
An EDM (or GR) experiment sensitive down to 
$\delta\!\Omega\approx1\,\mathrm{nrad/s}$ can be only constructed if this coherently 
accumulating field shape imperfection term can be controlled or cancelled. 
Controlling that contribution seems to be beyond experimental reach, since pushing down 
that term in a realistic setting would require suppressing $|B_{H}|$ 
below $\,{\approx}10^{-16}\,\mathrm{Tesla}$. A more promising approach would be 
a recent idea by R.~Talman \cite{talman2018}, called 
\emph{doubly-frozen spin storage ring}. The basic idea is that for certain 
particle type pairs (such as helion3 and proton beam pairs), it is possible 
to reach the frozen horizontal spin condition, i.e.\ Eq.(\ref{eqFr}), simultaneously 
in the same storage ring. Their spins will roll around the instantaneous 
beam-radial axis with rates $\delta\!\Omega_{1}$ and $\delta\!\Omega_{2}$, 
which both contain contributions from the conical magnetic field imperfection 
and from GR as dictated by Eq.(\ref{eqconical}), and eventually from EDM. 
The idea is to use an optimal weighted difference 
$\delta\!\Omega:=\delta\!\Omega_{1}-W\cdot\delta\!\Omega_{2}$ of the 
observables $\delta\!\Omega_{1}$ and $\delta\!\Omega_{2}$ of the two beams, 
such that the magnetic field conicality imperfection term, i.e.\ the contribution 
of the first term of Eq.(\ref{eqconical}), cancels. The optimal weighting 
factor $W$ is determined below.

\begin{figure}[!h]
\begin{center}
\includegraphics[width=9cm]{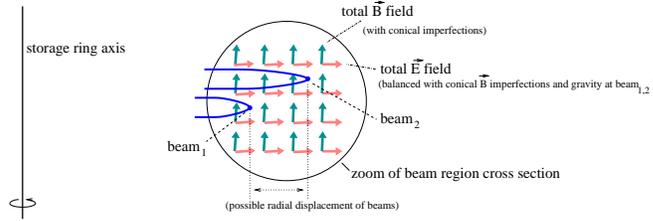}
\end{center}
\caption{(Color online) Illustration of the cross-section of an idealized 
doubly-frozen storage ring, proposed by R.~Talman \cite{talman2018}, with all 
the possible axially symmetric imperfections taken into account. Beams 
of two particle species are injected into the same storage ring fields, 
both under closed planar circular motion and frozen horizontal spin condition 
(their orbits might be possibly slightly displaced in terms of their bending 
radii). As such, the field imperfections felt by the two beams can be canceled to 
the first order, by combining their vertical polarization buildup rates 
$\delta\!\Omega_{1}$ and $\delta\!\Omega_{2}$.}
\label{figdoubly}
\end{figure}

An idealized planar circular doubly-frozen spin storage ring \cite{talman2018} 
assumes that beams of two particle species are stored in the same 
storage ring fields both in frozen horizontal spin condition. A cross section 
of such a setting is illustrated in Figure~\ref{figdoubly}. The parameters 
of ``$\mathrm{beam}_{i}$'' will be $\frac{m_{i}}{q_{i}}$ for mass-over-charge, 
$a_{i}$ for magnetic moment anomaly, $\beta_{i}\gamma_{i}$ for momentum-over-mass, 
and $L_{i}$ for bending radius ($i=1,2$ stands for the two beams). 
The idealized planar circular orbit of the two beams are not assumed to exactly coincide, 
that is why two possibly slightly different bending radii $L_{1,2}$ were assumed. 
The magnetic bending field is assumed to be (locally) homogeneous vertical, whereas the 
electrostatic bending field is assumed to be (locally) cylindrical, and therefore
\begin{eqnarray}
 B_{V}{}_{1}        & \;=\; & B_{V}{}_{2}        \;\quad\;=\; B_{V}    \quad\,=\; const, \cr
 L_{1}\,E_{H}{}_{1} & \;=\; & L_{2}\,E_{H}{}_{2} \;=\;        L\,E_{H} \;=\; const
\label{eqbend}
\end{eqnarray}
when sampled at the nominal planar circular orbit of the two beams. 
The tiny magnetic field conicality imperfection term (beam-radial component) 
is cylindrical, and needs to (locally) satisfy vacuum Maxwell equations, and therefore 
\begin{eqnarray}
 \Big. L_{1}\,B_{H}{}_{1} & \;=\; & L_{2}\,B_{H}{}_{2} \;=\;        L\,B_{H} \;=\; const
\label{eqbh}
\end{eqnarray}
holds (where $B_{H}$ is the amplitude of the idealized cylindrical stray 
fields, the main source of systematic errors in $\delta\!\Omega$). 
The amplitudes $E_{V}{}_{1}$ and $E_{V}{}_{2}$ exerted by the focusing 
quadrupoles are determined by the condition that the nominal 
orbits do not drift, i.e. the total electromagnetic and 
gravitational fields yield stationary orbits 
``$\mathrm{beam}_{1}$'' and ``$\mathrm{beam}_{2}$''. The above model takes 
into account all the possible instrumental imperfections which respect an 
idealized exact axial symmetry of the storage ring and beams. 
Therefore it provides an analytic means to quantify the most important 
systematic error contributions, which need to be dealt with, prior to 
investigation of further instrumental imperfection effects via detailed 
beamline simulations.

Given the identities for the fields Eq.(\ref{eqbend}) and Eq.(\ref{eqbh}), and 
that one would like to have the planar circular motion and frozen horizontal 
spin condition Eq.(\ref{eqFr}) is also required to hold, and the vertical 
forces are required to be also equilibrated, for both beams 
Eq.(\ref{eqconical}) will be satisfied. For the two beams, one gets
\begin{eqnarray}
 \Big. \delta\!\Omega_{1} & = & -\frac{q_{1}\,(1+a_{1})}{m_{1}}\frac{1}{\gamma_{1}{}^{2}}\,B_{H}{}_{1} \quad+\quad (-a_{1}\,\beta_{1}\gamma_{1}\;g/c) \cr
 \Big. \delta\!\Omega_{2} & = & -\frac{q_{2}\,(1+a_{2})}{m_{2}}\frac{1}{\gamma_{2}{}^{2}}\,B_{H}{}_{1}\frac{L_{1}}{L_{2}} \quad+\quad (-a_{2}\,\beta_{2}\gamma_{2}\;g/c)
\label{eqconical12}
\end{eqnarray}
for the polarization buildup signal by the magnetic field conicality imperfection 
$B_{H}{}_{1,2}$, and from the first order GR correction. 
From this equation, one can deduce the optimal weighting factor
\begin{eqnarray}
 W & = & \left(\frac{q_{1}\,(1+a_{1})}{m_{1}}\Big/\frac{q_{2}\,(1+a_{2})}{m_{2}}\right)\,\frac{\gamma_{2}{}^{2}}{\gamma_{1}{}^{2}}\frac{L_{2}}{L_{1}}
\label{eqw}
\end{eqnarray}
which is needed to be determined very accurately, in order to cancel the 
magnetic field conicality imperfection term from the combined signal.

Introduce the notation $A:=\frac{m}{q}\frac{G}{1+G}$. Moreover, assume that the particle pair properties
$\begin{array}{cccc}
 A_{1}/A_{2},\quad & G_{1}/G_{2},\quad & G_{2},\quad & L_{1}/L_{2} \cr
\end{array}$
can be accurately measured. As a function of these parameters, the nominal closed planar 
circular orbits ``$\mathrm{beam}_{1,2}$'' and their $(\beta\gamma)_{1,2}$ 
satisfying the frozen horizontal 
spin condition is uniquely determined, as illustrated in Figure~\ref{figdoubly}. 
Under that condition, the correction factor $W$, given by Eq.(\ref{eqw}) 
is also uniquely determined as a function of the above parameters, by a closed 
formula.

One can consider e.g.\ a set of particle species
\begin{eqnarray*}
\left\{\text{triton, helion3, proton, deuteron, 
$e^{+}$, $e^{-}$, C13 ion, F19 ion, $\mu^{+}$, $\mu^{-}$}\right\}
\end{eqnarray*}
for a doubly-frozen spin ring setting within the same ring ($\frac{L_{1}}{L_{2}}\approx1$). 
Using a Maple code one can numerically solve for the kinematic parameters of 
doubly-frozen spin configurations. Taking into account practical constraints, 
such as a reasonable bending radius $L\approx 8\,\mathrm{m}$ and a feasible 
electrostatic bending field $\vert E_{H}\vert\leq 8\,\mathrm{MV/m}$, it turns 
out that only the helion3--proton beam pairs are practically realistic.


For the pertinent helion3--proton doubly-frozen spin ring, the GR signals 
happen to combine constructively in the optimal weighted difference 
$\delta\!\Omega=\delta\!\Omega_{1}-W\cdot\delta\!\Omega_{2}$, and it is of the order 
of $-30\,\mathrm{nrad/s}$. It turns out that in order to cancel the magnetic field 
conicality imperfection in such setting, down to a factor of 10 signal-to-background ratio 
for the GR signal to be well detectable, the weighting factor $W$ needs to be 
known to such accuracy that $\vert\delta\!W\cdot B_{H}\vert\leq 10^{-17}\,\mathrm{Tesla}$ holds. 
The correction factor $W$ and its derivatives as a function of the particle parameters $A_{1}/A_{2}$, 
$G_{1}/G_{2}$, $G_{2}$ and the radial displacement parameter $L_{1}/L_{2}$ 
at the helion3--proton setting is of the order of $1$. It is not unrealistic 
to measure the particle parameters $A_{1}/A_{2}$, $G_{1}/G_{2}$, $G_{2}$ down 
to $10^{-10}$ relative accuracy. Moreover, with recent beam instrumentation technology, 
one can expect a measurement on the radial displacement $L_{1}/L_{2}$ down 
to a $10^{-7}$ relative accuracy. This means that the optimal weighting factor 
$W$ can be determined up to a systematic error of $\vert\delta\!W\vert\approx10^{-7}$. 
By means of the above error estimates, one would need to suppress the beam-radial 
stray field down to $\vert B_{H}\vert\approx 10^{-10}\,\mathrm{Tesla}$ for 
the GR signal (or an EDM signal) to be visible in a doubly-frozen spin ring setting.




\section{Concluding remarks}
\label{concludingremarks}

In this paper, a comparison of theoretical predictions \cite{orlov2012,obukov2016,laszlo2018} 
on GR systematics in frozen spin storage rings (EDM rings) \cite{semertzidis2016,talman2017} was presented. 
All theoretical predictions agree for magic momentum (that setting is 
only possible for beam particles with positive magnetic moment anomaly). 
The predictions \cite{obukov2016} and \cite{laszlo2018} differ substantially 
for negative magnetic moment particles (e.g.\ the experimentally rather 
important deuterons), or away from the magic momentum for positive 
magnetic moment particles. The difference is of the order of a factor of $2-8$. 
In this paper we showed that the pertinent difference comes from the assumption 
on the shape of the magnetic bending field. In \cite{obukov2016} an implicit 
assumption of an Earth-radial magnetic bending field was made, whereas in \cite{laszlo2018} 
the magnetic bending field was assumed to be asymptotically homogeneous (``vertical'' 
instead of Earth-radial). 
It was shown that, keeping in mind these assumptions, both results describe 
correct first order GR contributions to the EDM observable. 
The frozen spin storage ring experimental situation, however, is rather modeled 
by the assumptions of \cite{laszlo2018}, whereas the assumptions of 
\cite{obukov2016} rather fits to a special case of a so-called Koop spin wheel 
setting \cite{Mane2015} (i.e.\ when the spin rolls with a finite frequency, even without 
GR or EDM being present).

A further conclusion of the paper is that the conical shape imperfection 
of the magnetic bending field can be compensated to a first order by using 
the doubly-frozen spin storage ring setting proposed by Talman \cite{talman2018}. 
For instance for a helion3--proton doubly-frozen spin ring the 
GR systematics add constructively. If the radial displacements of the two 
beams can be measured down to $10^{-7}$ relative accuracy, one still needs 
to suppress the beam-radial component of the stray magnetic fields down 
to $10^{-10}\,\mathrm{Tesla}$, which is a significant experimental challenge.

\section*{Acknowledgments}

This work was supported in part by the Hungarian Scientific Research Fund 
(NKFIH 123842-123959).

\end{document}